\documentclass[aps,pra,showpacs,eqsecnum,twocolumn]{revtex4}
\usepackage{graphicx}
\usepackage{bm}
\begin{document}
\preprint{}
\title{Intrinsic metrological resolution as a distance
measure and nonclassical light}

\author{\'{A}ngel Rivas}
\email{A.Rivas@herts.ac.uk}
\affiliation{School of Physics, Astronomy and Mathematics,
University of Hertfordshire, College Lane, Hatfield, Herts,
AL10 9AB, United Kingdom}
\author{Alfredo Luis}
\email{alluis@fis.ucm.es}
\homepage{http://www.ucm.es/info/gioq}
\affiliation{Departamento de \'{O}ptica, Facultad de Ciencias
F\'{\i}sicas, Universidad Complutense, 28040 Madrid, Spain }
\date{\today}

\begin{abstract}
We elaborate on a Hilbert-Schmidt distance measure assessing
the intrinsic metrological accuracy in the detection of
signals imprinted on quantum probe states by signal-dependent
transformations. For small signals this leads to a
probe-transformation measure $\Lambda$ fully symmetric on the
probe $\rho$ and the generator $G$ of the transformation
$\Lambda (\rho, G) = \Lambda (G, \rho)$. Although $\Lambda$
can be regarded as a generalization of variance we show that
no uncertainty relation holds for the product of measures
corresponding to complementary generators. We show that all
states with resolution larger than coherent states are
nonclassical. We apply this formalism to feasible probes and
transformations.
\end{abstract}


\maketitle

\section{Introduction}

Quantum metrology is a key issue of quantum mechanics
involving fundamental concepts such as uncertainty
relations, complementarity, and nonclassical properties.
The basic objective of quantum metrology is to infer the
value of a given unknown signal $\chi$ as accurately as
possible at minimum cost. Signals $\chi$ are encoded on
quantum states by $\chi$-dependent unitary transformations
$U_\chi$ acting on an input probe state $\rho$, so that
all the information about $\chi$ is contained in the
output probe state $\rho_\chi = U_\chi \rho U^\dagger_\chi$.

The intrinsic accuracy of the detection depends primarily
on the contribution of two independent factors: the
preparation of the probe $\rho$ and the encoding
transformation $U_\chi$. While most works on quantum
metrology focus on the optimization of the input probe
state $\rho$ \cite{BM,noon}, recently it has been put
forward the advantages of investigating optimal encoding
transformations $U_\chi$ allowing more robust and accurate
detection schemes \cite{BHL}.

In this work we elaborate on the assessment of the intrinsic
accuracy by using the Hilbert-Schmidt distance between $\rho$
and $\rho_\chi$ \cite{LZ04,L03,DMMW00,HSexp,HSent}.
More specifically, we show that:

(i) For small signals the Hilbert-Schmidt distance becomes
a probe-transformation measure $\Lambda$ fully symmetric
on the probe $\rho$ and the generator $G$ of the
transformation $\Lambda (\rho, G) = \Lambda (G, \rho)$.
This acquires the form of a generalization of variance
previously used in quantum mechanics and classical optics
\cite{V04,PW01} (Sec. II).

(ii) We derive new expressions for the probe-generator
measure $\Lambda (\rho, G)$ (Sec. II).

(iii) We demonstrate that $\Lambda (\rho, G)$ is always
bounded from above by variance (Sec. II).

(iv) Despite resembling a generalization of variance,
we show that no uncertainty relation holds for the
product $\Lambda (\rho, A) \Lambda (\rho, B)$ for
complementary generators $A,B$ (Sec. II).

(v) We determine optimum generators leading to maximum
resolution for fixed input probe states.

(vi) We show that the probe-generator measure $\Lambda
(\rho, G)$ predicts nonclassical behavior (in the sense
of lack of positive phase-space representative $P(\alpha)$
\cite{WP}) for all states providing larger resolution
than coherent states (Sec. III).

(vii) We apply this formalism to feasible Gaussian
probes and standard transformations, determining
the probes reaching optimum intrinsic resolution
(Sec. IV).

\section{Intrinsic metrological resolution}

The signal to be detected $\chi$ is encoded in the input
probe state $\rho$ by a unitary transformation. For
definiteness, we focus on the most common and practical
case of constant generators $G$ independent of the
parameter $\chi$, $U_\chi = \exp \left ( i \chi G \right )$,
\begin{equation}
\rho_\chi = \exp \left ( i \chi G \right ) \rho
\exp \left ( - i \chi G \right ) ,
\end{equation}
where $G$ is the Hermitian generator of the transformation.
The intrinsic accuracy is given first and foremost by the
distinguishability between $\rho$ and $\rho_\chi$. A
convenient measure of distinguishability is the Hilbert-Schmidt
distance
\begin{equation}
d^2_{HS} (\chi) = \textrm{tr} \left [ \left ( \rho - \rho_\chi
\right )^2 \right ] = \textrm{tr} \left ( \rho^2 \right )
+  \textrm{tr} \left ( \rho^2_\chi \right ) - 2 \textrm{tr}
\left ( \rho \rho_\chi \right ) .
\end{equation}
The overlap term, $\textrm{tr} \left ( \rho \rho_\chi
\right )$, represents the survival probability, expressing
the inertia of $\rho$ to the changes generated by $G$.
There are many distance measures that may be used leading to
largely equivalent results. Among them, the Hilbert-Schmidt
distance is selected here because of conceptual and
computational simplicity, and its proximity with experimental
procedures \cite{HSexp}.

All similar measures of distinguishability between density
matrices $\rho$ and $\rho_\chi$ that may be used, such as
relative entropy, trace distance, Bures distance, Hellinger
distance, have drawbacks, such as not leading to proper
distance measures (relative entropy), lack of physical
interpretation or rather complex evaluation procedures (Hellinger
and Bures distances). Further comparison of properties can be
found in Ref. \cite{LZ04}. In this regard, the main drawback
of the Hilbert-Schmidt distance quoted in the literature is
the no-monotony decrease under quantum operations. As discussed
in Ref. \cite{HSent} this is a trouble to use this distance to quantify entanglement
when constructing entanglement monotones.
More properly, only monotonicity under local operations and
classical communication (LOCC) is needed, and general monotonicity
is a sufficient condition. Monotonicity is satisfied by
distinguishability measures such as relative entropy and Bures
distance \cite{HSent}.

Let us explain why monotonicity is not an essential property in our
context so that the Hilbert-Schmidt distance can be safely
used for our purposes. Monotonicity of a distance measure
$d(\rho_1 , \rho_2)$ means that $d(\rho_1 , \rho_2) \geq
d [\mathcal{E} (\rho_1), \mathcal{E}(\rho_2)]$ where
$\mathcal{E}$ are quantum operations (completely positive
trace-preserving maps) that include LOCC as a particular case.
The key point is that this property is always discussed in
terms of completely independent density matrices $\rho_{1,2}$.
However, in our case $\rho_{1,2}$ are not independent, since
one of them is the result of a signal-dependent transformation
$\mathcal{K}$ acting on the other $\rho_2 = \mathcal{K}
(\rho_1)$. Since in general $\mathcal{E} [\mathcal{K} (\rho)]
\neq \mathcal{K} [ \mathcal{E} (\rho )]$ there is no point
comparing $d [\rho , \mathcal{K} (\rho) ]$ and $d \{
\mathcal{E} (\rho) , \mathcal{E} [ \mathcal{K} (\rho) ]
\}$, since the last one is not of the form $d [ \rho^\prime ,
\mathcal{K} (\rho^\prime ) ]$. In physical terms $\mathcal{E}$
cannot act on $\rho$ and $\mathcal{K} (\rho)$ simultaneously,
since $\rho$ and $\mathcal{K}(\rho)$ never coexist (the former
precedes the latter). In other words, in our case the distance
$d$ must be understood as a function of $U_\chi$ and $\rho$
so that the application of other transformations is out of
the scope of our problem.

\subsection{Probe-generator measure}

Concerning metrological applications we are mostly
interested in very weak signals so we may consider the
limit $\chi \rightarrow 0$. Considering a power series
for $d^2_{HS} (\chi)$, the first nonvanishing term is
\begin{equation}
\label{loc}
d^2_{HS} (\chi) \simeq 2 \chi^2 \Lambda^2 \left ( \rho,G
\right ) ,
\end{equation}
where the probe-generator functional $\Lambda^2 \left (
\rho,G \right )$ is
\begin{equation}
\label{pgf}
\Lambda^2 \left ( \rho,G \right ) = \textrm{tr}
\left ( \rho^2 G^2 \right ) - \textrm{tr} \left ( \rho
G \rho G \right ) .
\end{equation}

Therefore $\Lambda \left ( \rho,G \right )$ measures
the capability of $G$ to efficiently imprint small signals
on the input probe $\rho$, so that larger $\Lambda^2
\left ( \rho, G \right )$ implies larger resolution.
The performance measure (\ref{loc}) can be regarded as a
generalization of the more familiar estimation uncertainty
$\delta \chi$ \cite{BM,noon}
\begin{equation}
\delta \chi \geq \frac{1}{2 \Delta_\rho G} ,
\end{equation}
where $\Delta_\rho G$ is the variance
\begin{equation}
\left ( \Delta_\rho G \right )^2 = \textrm{tr} \left (
\rho G^2 \right ) - \left [ \textrm{tr} \left ( \rho G
\right ) \right ]^2 .
\end{equation}

The probe-generator measure $\Lambda \left ( \rho,G \right )$
can be regarded as a generalization of variance since for
pure states $\rho^2 = \rho = | \psi \rangle \langle \psi |$
we have \cite{V04}
\begin{equation}
\label{ps}
\Lambda^2 \left ( | \psi \rangle ,G \right ) = \langle
\psi | G^2 | \psi \rangle - \langle \psi | G | \psi
\rangle^2 = \left ( \Delta_\psi G \right )^2 .
\end{equation}
However a literal strict interpretation of $\Lambda^2
\left ( \rho, G \right )$ as an uncertainty measure is
questionable, or even misleading, as shown below.

Let us note the complete symmetry between $\rho$ and $G$,
$\Lambda \left ( \rho,G \right )  = \Lambda \left ( G,
\rho \right )$. This symmetry suits to the idea of the
joint accountability of probe and transformation to
metrological performance.

\subsection{Equivalent expressions}

The probe-generator functional $\Lambda$ can be expressed
also as
\begin{equation}
\Lambda^2 \left ( \rho,G \right ) =  - \frac{1}{2}
\textrm{tr} \left ( [ \rho , G ]^2 \right ) = \frac{1}{2}
\textrm{tr} \left [ \left ( \frac{d \rho_\chi}{d \chi}
\right )^2_{\chi = 0} \right ] .
\end{equation}
This can be regarded as the analog of the Wigner-Yanase
skew information after replacing $\sqrt{\rho}$ by $\rho$
\cite{L03,LZ04}.

In terms of the spectrum and statistics of $G$ the following
expression holds
\begin{equation}
\label{sLG}
\Lambda^2 \left ( \rho,G \right ) = \frac{1}{2} \sum_{j,k}
\left ( g_j - g_k \right )^2 \left | \langle g_k | \rho |
g_j \rangle \right |^2 ,
\end{equation}
where $G | g_j \rangle = g_j | g_j \rangle$. This has
been used in classical optics to asses effective spatial
correlations of light beams \cite{PW01}.

According to the full symmetry between $G$ and $\rho$ we
can derive a relation dual to Eq. (\ref{sLG}) in terms of
the spectrum of $\rho$
\begin{equation}
\label{sr}
\Lambda^2 \left ( \rho,G \right ) = \frac{1}{2} \sum_{j,k}
\left ( r_j - r_k \right )^2 \left | \langle r_k | G | r_j
\rangle \right |^2 ,
\end{equation}
where $| r_j \rangle $ is the orthonormal basis  defined
by the eigenvectors of $\rho$, $\rho | r_j \rangle = r_j
| r_j \rangle$, including those with vanishing eigenvalue.

Furthermore, $\Lambda \left ( \rho , G \right )$ can be also
related with a kind of weighted version of variances of weak
values. This can be seen after expressing $\rho$ in the
$P(\alpha)$
representation
\begin{equation}
\label{P}
\rho = \int d^2 \alpha P(\alpha) | \alpha \rangle \langle
\alpha | ,
\end{equation}
where $| \alpha \rangle$ are coherent states \cite{WP}. Using
this representation in Eq. (\ref{pgf}) we get
\begin{eqnarray}
\label{w1}
\Lambda^2 \left ( \rho , G \right ) &=& \int d^2 \alpha d^2
\beta P(\alpha) P(\beta) \left | \langle \alpha | \beta
\rangle \right |^2\nonumber\\*
&&\times\left [ G^{(2)} \left ( \alpha , \beta
\right )  - \left | G^{(1)} \left ( \alpha , \beta \right )
\right |^2 \right ],
\end{eqnarray}
where $G^{(k)} \left ( \alpha , \beta \right )$ is the weak
value of $G^k$ in the coherent states $| \alpha \rangle$,
$| \beta \rangle$ \cite{weak}
\begin{equation}
\label{w2}
G^{(k)} \left ( \alpha , \beta \right ) = \frac{\langle
\beta | G^k | \alpha \rangle}{\langle \beta | \alpha \rangle}.
\end{equation}
Further expressions for $\Lambda^2 \left ( \rho , G \right )$
can be obtained for Cartesian conjugate variables in terms
of the Wigner function \cite{V04}.

\subsection{Variance bound}

Using Eq. (\ref{sLG}) we can demonstrate that $\Lambda$
is always bounded from above by variance
\begin{equation}
\label{bound}
\Lambda^2 \left ( \rho,G \right ) \leq \left ( \Delta_\rho
 G \right )^2 ,
\end{equation}
the equality being reached for pure states. To this end we
note that the density matrices $\rho$ and $\sigma$ have the
same variance $\Delta_\rho G = \Delta_\sigma G$, where
$\sigma$ is a pure state $\sigma = | \psi \rangle \langle
\psi |$ with
\begin{equation}
| \psi \rangle = \sum_k \sqrt{\langle g_k | \rho | g_k
\rangle} e^{i \varphi_k} | g_k \rangle ,
\end{equation}
where $\varphi_k$ are phases. Thus, by construction $\sigma$
is positive and Hermitian, and it holds
\begin{equation}
| \langle g_k | \sigma | g_j \rangle |^2 =
\langle g_k | \rho | g_k \rangle  \langle g_j | \rho |
g_j \rangle .
\end{equation}
The  equality $\Delta_\rho G = \Delta_\sigma G$ holds
because the variance of $G$ depends exclusively
on the diagonal matrix elements, that are equal for both
density matrices $\langle g_k | \rho | g_k \rangle = \langle
g_k | \sigma | g_k \rangle$. Furthermore, the positivity of
$\rho$ implies that
\begin{equation}
\label{rebound}
| \langle g_k | \rho | g_j \rangle |^2  \leq
\langle g_k | \rho | g_k \rangle  \langle g_j | \rho |
g_j \rangle = | \langle g_k | \sigma | g_j \rangle |^2 .
\end{equation}
Thus, from Eqs. (\ref{ps}), (\ref{sLG}), and (\ref{rebound})
we get
\begin{equation}
\Lambda^2 \left ( \rho,G \right ) \leq \Lambda^2 \left (
\sigma, G \right ) = \left ( \Delta_\sigma G \right )^2
= \left ( \Delta_\rho G \right )^2 ,
\end{equation}
which demonstrates Eq. (\ref{bound}).

\subsection{Lack of uncertainty relation}

The dependence of $\Lambda \left ( \rho,G \right )$ on the
coherence terms $\langle g_k | \rho | g_j \rangle$ reveals
that this is more than a measure of fluctuations. In this
regard, for example, we have $\Lambda \left ( \rho,G \right )
= 0$ if and only if
\begin{equation}
\rho = \sum_j p_j | g_j \rangle \langle g_j | ,
\end{equation}
with $p_j \geq 0$ and $\sum_j p_j =1$, so that
$\Delta_\rho G$ can take any value depending on $p_j$. This
is further discussed in Sec. IVC.

Moreover, despite that $\Lambda$ resembles a generalization
of variance, it does not lead to any uncertainty relation
when applied to complementary observables. More specifically
we show that there is no lower bound for the product $\Lambda
\left ( \rho , X \right ) \Lambda \left ( \rho , Y \right )$,
where $X$, $Y$ are two Cartesian conjugate observables,
analogous to position and linear momentum,
\begin{equation}
\label{XY}
X = \frac{1}{\sqrt{2}} \left ( a + a^\dagger \right ) ,
\qquad
Y = \frac{i}{\sqrt{2}} \left ( a^\dagger - a \right ) ,
\end{equation}
with $[X,Y] = i$, and $a^\dagger$, $a$ are creation and
annihilation operators with $[a,a^\dagger]=1$. In such
a case, we show in more detail in Sec. IVA that for the
squeezed vacuum states (\ref{sv}) we get
\begin{equation}
\Lambda^2 \left ( \rho , X \right ) =
\frac{1}{8 \Delta X (\Delta Y)^3} ,
\quad
\Lambda^2 \left ( \rho , Y \right ) =
\frac{1}{8 \Delta Y (\Delta X)^3} ,
\end{equation}
so that
\begin{equation}
\Lambda \left ( \rho , X \right ) \Lambda
\left ( \rho , Y \right ) =
\frac{1}{8 (\Delta X )^2 (\Delta Y)^2} .
\end{equation}
There is no lower bound for this product since $\Lambda
\left ( \rho , X \right ) \Lambda \left ( \rho , Y
\right ) \rightarrow 0$ when $\Delta X \rightarrow
\infty$ or $\Delta Y \rightarrow \infty$. Moreover, we
can notice that we can have $\Lambda \left ( \rho , X
\right ) \rightarrow 0$ and $\Lambda \left ( \rho , Y
\right ) \rightarrow 0$ simultaneously.

One may wonder whether the lack of uncertainty
relation is related with the lack of monotonicity of
the Hilbert-Schmidt distance. At first qualitative argument, we note that,
roughly speaking, lack of monotonicity is related with
increase of distance, while lack of uncertainty relation
is given by the fully opposite effect, i. e., decreasing
distances.  Anyway a simple and expeditious
procedure to solve this question is to show that the lack of
uncertainty relation also occurs for a probe-transformation
measure (quantum Fisher information) derived from a monotonic
distance measure \cite{LZ04,BC94,LU03}
\begin{equation}
\label{IF}
I_F (\rho,G) = \frac{1}{2} \sum_{j,k}
\frac{(r_j - r_k )^2}{r_j + r_k} \left | \langle r_j |
G | r_k \rangle \right |^2 ,
\end{equation}
where as in Eq. (\ref{sr}) $| r_j \rangle $ are the
eigenvectors of $\rho$ with eigenvalues $r_j$, and
the sum includes the cases with $r_j + r_k \neq 0$.
This measure is the infinitesimal local form of the Bures
distance \cite{LZ04,BC94}
\begin{equation}
d^2_B (\rho_1 , \rho_2) = 2 \left \{ 1 - \textrm{tr}
\left [ \left ( \rho_1^{1/2} \rho_2 \rho_1^{1/2}
\right )^{1/2} \right ] \right \} ,
\end{equation}
that fulfills monotonicity \cite{HSent}. Let us compute
$I_F (\rho, X)$ and $I_F (\rho,Y)$ for the quadratures
(\ref{XY}) in the thermal state
\begin{equation}
\label{ts}
\rho = (1-\xi) \sum_{n=0}^\infty \xi^n | n \rangle
\langle n |,
\end{equation}
where $\xi < 1$ is a real parameter, and $|n \rangle$ are
the number states $a^\dagger a |n \rangle = n|n \rangle$.
After a simple calculation it can be seen that
\begin{equation}
I_F (\rho, X) = I_F (\rho,Y) = \frac{1-\xi}{2(1
+ \xi)} .
\end{equation}
Therefore, when $\xi \rightarrow 1$ we get $I_F (\rho, X)
I_F (\rho,Y) \rightarrow 0$ demonstrating that the lack of
uncertainty relation is not a consequence of the lack of
monotonicity.

On the other hand, there are uncertainty relations involving
the product of $\Lambda ( \rho , A )$ for one observable
with a different measure $\tilde{\Lambda} ( \rho , B) \neq
\Lambda ( \rho , B)$ for the other one \cite{V04}, such as
\begin{equation}
\tilde{\Lambda}^2 \left ( \rho,G \right ) = \frac{1}{2}
\sum_{j,k} \left ( g_j + g_k \right )^2 \left | \langle
g_k | \rho | g_j \rangle \right |^2 .
\end{equation}
Related uncertainty relations has been proposed in
Refs. \cite{LZ04,LUO05}.

Seemingly Ref. \cite{PW01} introduces a lower bound for
the product of the same measure $\Lambda $ for Cartesian
conjugate variables within a classical optics framework.
However, a closer inspection reveals that such a bound is
actually another example of the unbalanced case in
Ref. \cite{V04} with different measures. Moreover, a
balanced uncertainty relation might be seemingly derived
from the following formula in Ref. \cite{L03}
\begin{equation}
\label{fuc}
4 I_W (\rho, A) I_W (\rho, B) \geq \left | \textrm{tr}
\left ( \rho [A,B] \right ) \right |^2 ,
\end{equation}
where $I_W (\rho, A) = \Lambda^2 ( \sqrt{\rho}, A )$ is the
Wigner-Yanase skew information which is the local infinitesimal
form of the Hellinger distance  \cite{LZ04,LU03}
\begin{equation}
d_H (\rho_1 , \rho_2) = \textrm{tr} \left [ \left ( \rho_1^{1/2}
- \rho_2^{1/2} \right )^2 \right ] .
\end{equation}
However, Eq. (\ref{fuc}) does not hold \cite{fn1}, as revealed
by a simple counterexample since Eq. (\ref{fuc}) is violated
by
\begin{equation}
\rho = \pmatrix{0.75 & 0 \cr 0 & 0.25} ,
\end{equation}
for $A=\sigma_x$ and $B=\sigma_y$, where $\sigma_{x,y}$
are the corresponding Pauli matrices. In this case we have
$I_W (\rho ,\sigma_x) = I_W (\rho ,\sigma_y) = 0.134$ while
$| \textrm{tr} ( \rho [\sigma_x,\sigma_y ] ) |^2 =1$. The
lack of a true uncertainty relation for this measure can be
further demonstrated again by computing $I_W (\rho, X)$ and $I_W
(\rho,Y)$ for the quadratures (\ref{XY}) in the thermal state
(\ref{ts}), leading to
\begin{equation}
I_W (\rho, X) = I_W (\rho,Y) = \frac{1-\sqrt{\xi}}{2( 1 +
\sqrt{\xi})} ,
\end{equation}
so that $I_W (\rho, X) I_W (\rho,Y) \rightarrow 0$ when $\xi
\rightarrow 1$. This agrees with the general relation
$I_W (\rho, A) \leq I_F (\rho, A)$ \cite{LU03}.

Finally, it is worth noting that there are fluctuation
measures that seem to defy the existence of an uncertainty
relation for complementary observables, as shown in
Ref. \cite{ZPV}. Nevertheless, note that $\Lambda \rightarrow 0$ does
not mean in this context arbitrary precision. On the
contrary, it means complete lack of measuring resolution
in the form of indistinguishability between input $\rho$
and transformed $\rho_\chi$ states. We recall that as shown in
this work $\Lambda$ is a measure of intrinsic resolution
rather than a measure of uncertainty.

\subsection{Optimum generators}

The full symmetry between states and transformations
invites to look for the optimum generator $G$ leading
to maximum intrinsic resolution for fixed probe state
$\rho$. This is the dual of the most common operation
in quantum metrology of determining the optimum $\rho$
for fixed $G$ \cite{BM}.

Despite the symmetry between $\rho$ and $G$ these two
operators belong to different classes, unit-trace
Hermitian positive definite for $\rho$, and just Hermitian
for $G$. Therefore, in order to fully exploit the $\rho$,
$G$ symmetry in the above calculus we restrict ourselves
to finite-trace Hermitian positive definite generators.
This implies no loss of generality for finite-dimensional
$G$ or $\rho$, since finite dimension guarantees finite
trace and also positivity by adding a constant to $G$
without altering neither $U$ nor $\Lambda$.

In such a case $\rho$ and $G$ can safely exchange their
roles in the above calculus so that the variance bound
(\ref{bound}) leads us to consider \emph{pure} generators
of the form $G \propto | \psi \rangle \langle \psi |$
where $| \psi \rangle$ is a normalized vector to be
determined by the condition of maximum variance of the
$\rho$ operator
\begin{equation}
\left ( \Delta_\psi \rho \right )^2 = \langle \psi |
\rho^2 | \psi \rangle - \langle \psi | \rho | \psi
\rangle^2 .
\end{equation}
Maximum variance is given by the extremal dichotomic
statistics provided by states of the form
\begin{equation}
| \psi \rangle = \frac{1}{\sqrt{2}} \left (
| r_{\textrm{max}} \rangle +
| r_{\textrm{min}} \rangle \right ),
\end{equation}
where $| r_{\textrm{max,min}} \rangle$ are the eigenvectors
of $\rho$ with extreme eigenvalue. This is a coherent
superposition of states with extremes eigenvalues fully
analogous to an equivalent result for probe optimization
for fixed generator \cite{noon}. Maybe generators of the
form  $G \propto | \psi \rangle \langle \psi |$ are
rather exotic and void of practical implementations.
Nevertheless this example illustrates the fundamental
symmetric role of probes and generators.

\section{Optimum nonclassical states}

In this section we show that for three representative
generators $G$ (position, number, and number difference)
all states providing larger resolution than coherent states
are nonclassical, in the sense of lack of positive
definite $P(\alpha)$ distribution in Eq. (\ref{P}).

\subsection{Position operator}

Let us consider transformations generated by the position
operator $X$ in Eq. (\ref{XY}) that produces the displacement
of the conjugate observable $U_\chi^\dagger Y U_\chi =
Y + \chi$. Better resolution than the one provided by
coherent states $| \alpha \rangle$ (with $a | \alpha
\rangle = \alpha | \alpha \rangle$) means that
\begin{equation}
\label{FLX}
\Lambda^2 \left ( \rho , X \right ) > \Lambda^2 \left (
| \alpha \rangle , X \right ) = \left ( \Delta_\alpha X
\right )^2 = \frac{1}{2} .
\end{equation}
By using Eqs. (\ref{w1}) and (\ref{w2}) we get
\begin{eqnarray}
&\Lambda^2 \left ( \rho , X \right )& = \frac{1}{2} \int d^2
\alpha d^2 \beta P(\alpha) P ( \beta ) \left | \langle
\alpha | \beta \rangle \right |^2 \nonumber\\*
&&\times\left \{ 1 + \textrm{Re}
\left [ \left ( \alpha + \beta^\ast \right )^2 \right ]
- \left | \alpha + \beta^\ast \right |^2 \right \} ,
\end{eqnarray}
where $\textrm{Re}$ represents the real part. We have used
that $\Lambda^2 \left ( \rho , X \right )$ is a real quantity,
so that the contribution from the imaginary part of $( \alpha
+ \beta^\ast )^2$ must vanish.

Condition (\ref{FLX}) is equivalent to
\begin{equation}
\label{ccon}
\int d^2 \alpha d^2 \beta P(\alpha) P ( \beta )
f ( \alpha , \beta ) < 0,
\end{equation}
where
\begin{equation}
f ( \alpha , \beta ) = 1-  \left | \langle
\alpha | \beta \rangle \right |^2 \left \{ 1 + \textrm{Re}
\left [ \left ( \alpha + \beta^\ast \right )^2 \right ]
- \left | \alpha + \beta^\ast \right |^2 \right \} ,
\end{equation}
and we have used that $\int d^2 \alpha P(\alpha) =1$.
Since for all complex numbers $A$ it holds $\textrm{Re} A
\leq |A|$ we get
\begin{equation}
1 + \textrm{Re}
\left [ \left ( \alpha + \beta^\ast \right )^2 \right ]
- \left | \alpha + \beta^\ast \right |^2 \leq 1 ,
\end{equation}
so that $f ( \alpha , \beta ) \geq 0$ and Eq. (\ref{ccon})
implies $P(\alpha) <0$ for some $\alpha$. Therefore, the
improvement of the intrinsic metrological resolution
beyond coherent states implies nonclassical character for
the probe.

\subsection{Number operator}

Next we consider transformations generated by the number
operator $G = N =a^\dagger a$, so that $\chi$ is a phase
shift $U_\chi^\dagger a U_\chi = \exp (i \chi ) a$.
Intrinsic resolution beyond coherent states $| \alpha
\rangle$ means
\begin{equation}
\label{LN0}
\Lambda^2 \left ( \rho , N \right ) > \Lambda^2 \left (
| \alpha \rangle , N \right ) = \left ( \Delta_\alpha N
\right )^2 = \langle \alpha | N | \alpha \rangle.
\end{equation}
In these schemes the accuracy increases when the mean number
increases. Therefore, for a proper comparison between the
performances provided by different states it is convenient
to consider probes with fixed mean number $\textrm{tr}
\left ( N \rho \right ) = \langle \alpha | N | \alpha
\rangle$, so that condition (\ref{LN0}) becomes
\begin{equation}
\label{LN}
\Lambda^2 \left ( \rho , N \right ) > \textrm{tr} \left (
N \rho \right ) .
\end{equation}
Using the $P ( \alpha )$ representation we get
\begin{eqnarray}
\label{LNP}
&\Lambda^2&  \left(\rho , N \right ) = \left(\int d^2
\alpha d^2 \beta P(\alpha) P ( \beta ) \left | \langle
\alpha | \beta \rangle \right |^2 \right.  \nonumber\\*
&&\times \left. \left\{ \textrm{Re}
\left [ \left ( \alpha \beta^\ast \right )^2 \right ]
- \left | \alpha \beta^\ast \right |^2 \right \}\right)+\textrm{tr} \left ( N \rho^2 \right ),
\end{eqnarray}
where again $\textrm{Re}$ represents the real part and
we have used that $\Lambda^2 \left ( \rho , N \right )$
is a real quantity so that the imaginary part of $( \alpha
\beta^\ast )^2$ does not contribute. Since for all $\rho$
it holds $\textrm{tr} ( N \rho^2 ) \leq \textrm{tr} ( N
\rho )$ condition (\ref{LN}) implies
\begin{equation}
\int d^2 \alpha d^2 \beta P(\alpha) P ( \beta ) \left |
\langle \alpha | \beta \rangle \right |^2 \left \{
\textrm{Re} \left [ \left ( \alpha \beta^\ast \right )^2
\right ] - \left | \alpha \beta^\ast \right |^2 \right \}
>0 .
\end{equation}
Taking into account that $\textrm{Re} A \leq |A|$ we get
\begin{equation}
\textrm{Re} \left [ \left ( \alpha \beta^\ast \right )^2
\right ] - \left | \alpha \beta^\ast \right |^2 \leq 0,
\end{equation}
so that the fulfillment of condition (\ref{LN}) requires
nonclassical probes.

\subsection{Number difference operator}

Finally we consider a two-mode situation with generator
$ G = J_z = N_1 - N_2$ with $N_j = a_j^\dagger a_j$,
$j=1,2$, so that in this case $\chi$ is a phase-difference
shift $U_\chi^\dagger a_1 a_2^\dagger U_\chi = \exp (2 i
\chi ) a_1 a_2^\dagger$. This is perhaps the most common
transformation in quantum metrology including all linear
interferometric and spectroscopic schemes. Intrinsic
resolution beyond the one provided by two-mode coherent
states $| \alpha_1 \rangle | \alpha_2 \rangle$ means that
\begin{equation}
\label{LJ}
\Lambda^2 \left ( \rho , J_z \right ) > \left (
\Delta_{\alpha_1,\alpha_2} J_z \right )^2 = \textrm{tr}
\left [ \left ( N_1 + N_2 \right ) \rho \right ],
\end{equation}
and here again we consider the same total mean number
$N_1 + N_2$ in $\rho$ and $| \alpha_1 \rangle | \alpha_2
\rangle$. In this case
\begin{eqnarray}
&\Lambda^2 \left ( \rho , J_z \right )& = \Lambda^2
\left ( \rho , N_1 \right ) + \Lambda^2 \left ( \rho ,
N_2 \right ) \nonumber \\*
&&- 2 \left [ \textrm{tr} \left ( \rho^2 N_1
N_2 \right ) - \textrm{tr} \left ( \rho N_1 \rho N_2
\right ) \right ] .
\end{eqnarray}
Denoting by $\bm{\alpha} = (\alpha_1 , \alpha_2 )$ and
$\bm{\beta} = (\beta_1 , \beta_2 )$ the last term
enclosed within square brackets can be rewritten as
\begin{equation}
\int d^2 \bm{\alpha} d^2 \bm{\beta} P(\bm{\alpha})
P( \bm{\beta}) \left | \langle \bm{\alpha} | \bm{\beta}
\rangle \right |^2 f ( \bm{\alpha} , \bm{\beta} ) ,
\end{equation}
where
\begin{eqnarray}
f ( \alpha , \beta ) &=& \textrm{Re} \left ( \beta_1^\ast
\alpha_1 \beta_2^\ast \alpha_2 - \beta_1 \alpha_1^\ast
\beta_2^\ast  \alpha_2 \right )= \nonumber\\
&& - 2 \textrm{Im} \left (
\beta^\ast_1 \alpha_1 \right ) \textrm{Im} \left (
\beta^\ast_2 \alpha_2 \right ) ,
\end{eqnarray}
so that Eqs. (\ref{LNP}) and (\ref{LJ}) imply
\begin{equation}
\Lambda^2 \left ( \rho , J_z \right ) =
\int d^2 \bm{\alpha} d^2 \bm{\beta} P(\bm{\alpha})
P( \bm{\beta}) \left | \langle \bm{\alpha} | \bm{\beta}
\rangle \right |^2 h ( \bm{\alpha} , \bm{\beta} ) > 0,
\end{equation}
where
\begin{eqnarray}
h ( \bm{\alpha} , \bm{\beta} ) &=&
\textrm{Re}
\left [ \left ( \alpha_1 \beta_1^\ast \right )^2 \right ]
- \left | \alpha_1 \beta_1^\ast \right |^2 + \textrm{Re}
\left [ \left ( \alpha_2 \beta_2^\ast \right )^2 \right ]
 \nonumber\\*
&-& \left | \alpha_2 \beta_2^\ast \right |^2 + 4 \textrm{Im} \left ( \beta^\ast_1 \alpha_1 \right )
\textrm{Im} \left ( \beta^\ast_2 \alpha_2 \right ) .
\end{eqnarray}
By expressing $\alpha_j \beta_j^\ast$ in terms of their
real and imaginary parts, $\alpha_j \beta_j^\ast = a_j +
i b_j$, we get that $h ( \bm{\alpha} , \bm{\beta} )$ is
always negative or zero
\begin{equation}
h ( \bm{\alpha} , \bm{\beta} ) =
- 2 \left ( b_1 - b_2  \right )^2 \leq 0 .
\end{equation}
Therefore, also for this two-mode generator, resolution
beyond coherent states implies nonclassical behavior.

\section{Examples}

In this section we apply the above formalism to practical
probes with Gaussian wave-function (coherent, squeezed,
and thermal-chaotic states), and standard generators
(phase-space displacements and rotations) looking for
optimal intrinsic resolution. Focusing on light beams we
consider definite energy resources represented by fixed
mean number of photons. Finally we present also the simplest
case of a two-dimensional system in order to illustrate
some properties of $\Lambda$.

\subsection{Optimum Gaussian states for displacements}

Let us consider signals encoded by $Y$-displacements
generated by $X$. Our objective is to obtain maximum
intrinsic resolution when using probe Gaussian states
with constant energy, i. e., fixed mean number of photons
$n$,
\begin{equation}
n = \textrm{tr} \left ( \rho a^\dagger a \right ) =
\frac{1}{2} \left [ \textrm{tr} \left ( \rho X^2 \right )
+ \textrm{tr} \left ( \rho Y^2 \right ) - 1 \right ] ,
\end{equation}
which is equivalent to
\begin{equation}
\label{mn}
n = \textrm{tr} \left ( \rho a^\dagger a \right ) =
\frac{1}{2} \left [ ( \Delta X )^2 + ( \Delta Y )^2
+ \langle X \rangle^2 + \langle Y \rangle^2 - 1 \right ] ,
\end{equation}
where $\langle X \rangle$, $\langle Y \rangle$, $\Delta X$,
$\Delta Y$, are the corresponding mean values and variances,
with $\Delta X \Delta Y \geq 1/2$.

Calculations are much simplified if we use the Wigner
representation so that the overlap between two states
$\rho$, $\rho^\prime$ is computed as \cite{WP}
\begin{equation}
\label{pi1}
\textrm{tr} \left ( \rho \rho^\prime \right ) = 2 \pi
\int dx dy W_\rho \left ( x, y \right ) W_{\rho^\prime}
\left ( x, y \right ).
\end{equation}
Under phase-space displacements the  Wigner function
transforms as a classical distribution
\begin{equation}
\label{pi2}
W_{\rho_\chi} \left ( x, y + \chi \right ) = W_\rho
\left ( x, y \right ) .
\end{equation}
Furthermore, Wigner functions for Gaussian states are
always positive.

Let us consider $\rho$ with Gaussian Wigner function
\begin{equation}
\label{sv}
W_\rho \left ( x, y \right ) = \frac{1}{2 \pi \Delta X
\Delta Y} \exp \left [ - \frac{(x-\langle X \rangle )^2}
{2 (\Delta X)^2} - \frac{(y - \langle Y \rangle)^2}{2
(\Delta Y)^2} \right ] .
\end{equation}
The Hilbert-Schmidt distance between $\rho_\chi$ and $\rho$
for the state (\ref{sv}) can be computed exactly
\begin{equation}
\label{Dchi}
d^2_{HS} (\chi) = \frac{1}{\Delta X \Delta Y} \left \{ 1 -
\exp \left [ - \frac{\chi^2}{4 (\Delta Y)^2}
\right ] \right \} .
\end{equation}
Optimum resolution requires minimum $\Delta Y$ for fixed
$n$. As reflected by Eq. (\ref{mn}) displacement and
squeezing compete for the photons since reducing $\Delta Y$
implies increasing $\Delta X$. Since displacement has no
effect on the resolution (\ref{Dchi}) optimum results are
obtained by employing all photons in squeezing so that
$\langle X \rangle = \langle Y \rangle = 0$.

For small signals Eq. (\ref{Dchi}) becomes
\begin{equation}
d^2_{HS} (\chi) \simeq \frac{\chi^2}{4 \Delta X
(\Delta Y)^3} ,
\end{equation}
so that
\begin{equation}
\label{LX}
\Lambda^2 \left ( \rho , X \right ) =
\frac{1}{8 \Delta X (\Delta Y)^3} ,
\end{equation}
and similarly
\begin{equation}
\label{LY}
\Lambda^2 \left ( \rho , Y \right ) =
\frac{1}{8 \Delta Y (\Delta X)^3} .
\end{equation}

For coherent probes $\Delta X = \Delta Y = 1 /\sqrt{2}$
we get $\Lambda^2 \left ( | \alpha \rangle, X \right ) =
1/2$. Looking for larger resolutions let us express
$\Delta X $ and $\Delta Y$ as functions of the mean
energy $n$ and the factor $p = \Delta X
\Delta Y \geq 1/2$
\begin{eqnarray}
\label{LXLY}
& ( \Delta X )^2 \simeq n + \sqrt{n^2 - p^2} , &
\nonumber \\
& ( \Delta Y )^2 \simeq n - \sqrt{n^2 - p^2} , &
\end{eqnarray}
where we have considered $2n + 1 \simeq 2n$. Let us note
that $p$ represents the purity of the probe since
$\textrm{tr} (\rho^2 ) = 1/(2p)$. On the other hand,
since $\langle X \rangle = \langle Y \rangle =0$ and
$( \Delta Y )^2 \simeq p^2/(2n)$ for $n>>1$, we have
that $n$ represents the amount of squeezing for fixed $p$.
Using Eq. (\ref{LXLY}) in Eq. (\ref{LX}) we get
\begin{equation}
\Lambda^2 \left ( \rho , X \right ) \simeq
\frac{1}{8 p \left ( n - \sqrt{n^2 - p^2} \right )} .
\end{equation}
Maximum metrological resolution implies maximum $\Lambda$
which is achieved for minimum $p$, i. e., $p=1/2$ (pure
minimum uncertainty states), so that for $n>>1$
\begin{equation}
\Lambda^2 \left ( \rho , X \right )
\simeq 2 n.
\end{equation}
Therefore the maximum accuracy for displacements of
Gaussian probes is obtained for pure nonclassical
squeezed vacuum states with
\begin{equation}
\label{ss}
\left ( \Delta Y \right )^2 \simeq \frac{1}{8 n},
\quad
\left ( \Delta X \right )^2 \simeq 2 n .
\end{equation}

\subsection{Optimum Gaussian states for phase shifts}

In this case we consider signals encoded on Gaussian
states with fixed mean energy (mean number $n$) by
phase-shifts generated by the number operator $N =
a^\dagger a$. The goal is to obtain maximum intrinsic
resolution. Here again we use Gaussian Wigner functions
centered at point $(x_0,0)$ and squeezed along the $y$
direction, so that the overlap under small rotations is
minimum,
\begin{equation}
W_\rho \left ( x, y \right ) = \frac{1}{2 \pi
\Delta X \Delta Y} \exp \left [ - \frac{(x-x_0)^2}{2
(\Delta X)^2} - \frac{y^2}{2 (\Delta Y)^2} \right ] ,
\end{equation}
with $\Delta X > \Delta Y$ and
\begin{equation}
\label{omn}
n = \textrm{tr} \left ( \rho a^\dagger a \right ) =
\frac{1}{2} \left [ ( \Delta X )^2 + ( \Delta Y)^2
+ x_0^2 - 1 \right ] .
\end{equation}
Also in this case the Wigner function transforms just
by the transformation of its variables, as a classical
distribution
\begin{equation}
W_{\rho_\chi} (x,y) =  W_{\rho} \left ( x \cos \chi
+ y \sin \chi, y \cos \chi - x \sin \chi \right ) .
\end{equation}

All this leads to the following exact result for the
Hilbert-Schmidt distance between $\rho$ and $\rho_\chi$
\begin{equation}
\label{DAB}
d^2_{HS} (\chi) = \frac{1}{\Delta X \Delta Y} \left ( 1 -
\frac{\exp A}{\sqrt{B}} \right ) ,
\end{equation}
with
\begin{equation}
A = - \frac{x_0^2 \sin^2 (\chi/2)}{(\Delta X)^2
\sin^2 (\chi /2) + (\Delta Y)^2 \cos^2 (\chi /2)} ,
\end{equation}
and
\begin{eqnarray}
B &=& \frac{1}{8 (\Delta X)^2 (\Delta Y)^2} \left\{(\Delta X)^4 + (\Delta Y)^4 + 6 (\Delta X)^2(\Delta Y)^2 \right. \nonumber\\*
&-& \left. \left [ (\Delta X)^2 - (\Delta Y)^2
\right ]^2 \cos (2 \chi )\right\} .
\end{eqnarray}
From Eqs. (\ref{loc}) and (\ref{DAB}) we get
\begin{equation}
\Lambda^2 \left (\rho , N \right ) = \frac{\left [
( \Delta X )^2 - ( \Delta Y )^2 \right ]^2 + 2 x_0^2 (
\Delta X )^2}{16 ( \Delta X )^2 ( \Delta Y )^2} .
\end{equation}
From Eq. (\ref{omn}), denoting again $p = \Delta X
\Delta Y$, we have
\begin{equation}
\left [ ( \Delta X )^2 - ( \Delta Y )^2 \right ]^2 =
( 2 n + 1 - x_0^2 )^2 - 4 p^2 ,
\end{equation}
and
\begin{equation}
( \Delta X )^2 = \frac{1}{2} \left [ 2 n + 1 - x_0^2
+ \sqrt{ (2 n + 1 - x_0^2 )^2 - 4 p^2} \right ] .
\end{equation}
Therefore for fixed $n$ and $x_0$ the maximum $\Lambda$
is obtained for minimum $p$, i. e., for pure states
$p = 1/2$. Then, the maximum when $x_0$ is varied for
fixed $n$ is obtained when $x_0 =0$, leading to
\begin{equation}
\label{rncs}
\Lambda^2 \left (\rho , N \right ) \simeq n^2 ,
\end{equation}
which is achieved for the same squeezed vacuum states in
Eq. (\ref{ss}).  The resolution (\ref{rncs}) is notably
larger than the one obtained for coherent states
$\Delta X = \Delta Y = 1/\sqrt{2}$ with the same mean
number $n = x_0^2 /2$
\begin{equation}
\Lambda^2 \left (| \alpha \rangle , N \right )
\simeq \frac{n}{2}.
\end{equation}

\subsection{Two-dimensional space}

Let us consider an arbitrary Hermitian generator in
a two-dimensional space, which can be always expressed
in the basis of its eigenvectors as
\begin{equation}
G = \pmatrix{ g_1 & 0 \cr 0 & g_2} .
\end{equation}
An arbitrary probe state reads in the same basis
\begin{equation}
\rho = \pmatrix{ q & \mu \sqrt{q (1-q)} \cr
\mu^\ast \sqrt{q (1- q )} & 1-q} ,
\end{equation}
with $1 \geq  q \geq 0$ and $|\mu | \leq 1$. In this case
we have
\begin{equation}
\Lambda^2 (\rho , G) =
q(1-q) (g_1 - g_2 )^2 |\mu |^2 =|\mu |^2
( \Delta_\rho  G)^2 ,
\end{equation}
which agrees with the general bound in Eq. (\ref{bound}).

We can appreciate that the main difference between
$\Lambda (\rho, G)$ and $\Delta_\rho G$ is that $\Lambda $
depends on the coherence term $\mu \propto \langle g_1
| \rho | g_2 \rangle $. In particular, for $\mu =0$ we
have $\Lambda (\rho, G) = 0$ irrespectively of
$\Delta_\rho G$. This is because in such a case $[\rho,G]
=0 $ so that $\rho$ is invariant under the transformations
generated by $G$. Therefore, although for pure states
$\Lambda$ becomes variance, in the general case it is
significantly different from a measure of fluctuations.

\section{Conclusions}

The Hilbert-Schmidt distance is a simple measure of the
intrinsic metrological resolution provided by a combination
of initial probe state and imprinting transformation. We
have shown that for small signals this becomes a
probe-transformation measure fully symmetricon the input
probe state and on the generator of the transformation.
For pure states this coincides with variance, but in the
general case it expresses a rather different concept, i. e.,
metrological resolution. The idea that the probe-transformation
measure is not always a proper measure of uncertainty is
demonstrated by the lack of an uncertainty relation when
applying it to complementary generators. We have shown that
this is not related with the lack of monotonicity of the
Hilbert-Schmidt distance, since the lack of uncertainty
relation is reproduced as well by monotonic measures.

Furthermore, we have shown that all states providing
resolution larger than coherent states are nonclassical.
This is remarkable since this corresponds to states
with probe-transformation measure larger than for
coherent states, while nonclassical behavior is usually
ascribed to reduced variances. Nevertheless, for the
examples presented in Sec. IV, probe-transformation
measure beyond coherent states becomes fully equivalent
to reduced variance below the vacuum level.

\section*{Acknowledgment}

A.R. acknowledges financial support from the University
of Hertfordshire and the EU Integrated Project QAP. A.L.
acknowledges support from the Universidad Complutense
Project No. PR1-A/07-15378.


\begin{thebibliography}{00}
\bibitem{fn1}
This is essentially because Eq. (\ref{fuc}) is derived in
Ref. \cite{L03} from a nonpositive definite inner product.
\end{thebibliography}
\end{document}